\documentstyle{l-aa}

\def\Lsol{\,L_\odot}
\def\kms{kms$^{-1}$}
\def\vt{$v_{\rm t}$}

\def\teff{$T_{\rm eff}$~}
\def\C12C13{$^{12}$C/$^{13}$C}

\def\sixt`enth{\textstyle{1\over{16}}}
\begin{document}
\input{psfig.tex}

\thesaurus{ 7(08.01.1 - 02.14.1 - 10.05.1) }
\title{ Abundances of light elements in metal-poor stars. III. }
\subtitle{ Data analysis and results }
\author{ E. Carretta$^1$, R.~G. Gratton$^1$, C. Sneden$^2$ } 
\offprints{ E. Carretta }
\institute{ 
$^1$Osservatorio Astronomico di Padova, Vicolo dell'Osservatorio, 5,
    I-35122 Padova, ITALY\\
$^2$University of Texas at Austin and McDonald Observatory, U.S.A.}
\date{}

\maketitle
\markboth{E.~Carretta et al.}{}

\begin{abstract}

We present the results of the analysis of an extensive set of new and
literature high quality data concerning Fe, C, N, O, Na, and Mg. This analysis
exploited the \teff\ scale determined in Gratton et al. (1996a), and
the non-LTE abundance corrections computed in Gratton et al. (1999a). Results
obtained with various abundance indices are discussed and compared. Detailed
comparison with models of galactic chemical evolution will be presented in
future papers of this series.

Our non-LTE analysis yields the same O abundances from both permitted and
forbidden lines for stars with \teff$>$4600~K, in agreement with King (1993),
but not with other studies using a lower \teff-scale for subdwarfs. However, we
obtain slightly smaller O abundances for the most luminous metal-poor field
stars than for fainter stars of similar metallicities, an effect attributed to
inadequacies of the adopted model atmospheres (Kurucz 1992, with overshooting)
for cool stars. We find a nearly constant O overundance in metal-poor stars
([Fe/H]$<-0.8$), at a mean value of $0.46\pm 0.02$~dex ($\sigma=0.12$, 32
stars), with only a gentle slope with [Fe/H] ($\sim -0.1$); this result is
different from the steeper slope recently obtained using OH band in the near
UV.

If only {\it bonafide} unmixed stars are considered, C abundances scale with Fe
ones (i.e. [C/Fe]$\approx 0$) down to [Fe/H]$\sim -2.5$. Due to our adoption
of a different \teff\ scale, we do not confirm the slight C excess in the most
metal poor disk dwarfs ($-0.8<$[Fe/H]$<-0.4$) found in previous investigations.

Na abundances scale as Fe ones in the high metallicity regime, while metal-poor
stars present a Na underabundance. None of the field stars analyzed belong to
the group of O-poor and Na-rich stars observed in globular clusters. 
Na is deficient with respect to Mg in halo and thick disk stars; within these
populations, Na deficiency may be a slow function of [Mg/H]. Solar [Na/Mg]
ratios are obtained for thin disk stars.

\end{abstract}

\section {Introduction}

The abundances of light elements (CNO, Na and Mg) in metal-poor stars provide
basic constraints for models of the chemical evolution of our Galaxy. These
elements are produced by different mechanisms in various astronomical sites.
The ejecta of core collapse SNe are expected to be very rich in O, while they
are expected to contribute only a fraction of the present C content of
the interstellar medium (ISM) (see e.g. Woosley \& Weaver 1986, 1995; 
Timmes et al. 1995; Thielemann et al 1990). On the other hand, mixing and/or
severe mass loss may bring to the surface large amounts of freshly synthesized
C from the outer and cooler regions of He-burning shells; this C may then be
returned to the ISM by stellar winds (Iben \& Renzini 1983 and references
therein). Since these are not fully understood mechanisms, and they may be
active in stars of very different masses (and hence lifetimes), the run of the
[C/Fe] ratio with time or overall metal abundance [Fe/H] \footnote{ In this
paper we will use two different notations for abundances: log~n(A) is the
abundance (by number) of the element A in the usual scale where log~n(H)=12;
[A/H] is the logarithmic ratio of the abundances of elements A and H in the
star, minus the same quantity in the Sun.} is presently not well constrained
by stellar and galactic evolution models. Much more insight can be gained from
observations.

Even more intriguing is the case for N, since this element is not directly
produced in large amounts by hydrostatic He-burning; N synthesis requires
recycling of material from a region where H-burning occurs through the CNO
cycle. In the case of massive stars, N synthesis is then expected to be a
secondary process (see e.g. Woosley and Weaver 1995), although under certain
circumstances even a primary behaviour could be obtained (Timmes et al. 1995).
Observational data indicate that large amounts of N are present in the outer
layers of evolved stars over a wide mass range. From there, N can be lost to
the ISM through quiescent or violent stellar winds, as e.g. in the case of
planetary nebulae. This N might result from the processing of either the
original C (leading to a secondary-like behaviour), or of freshly synthesized
$^{12}$C as a consequence of CN-cycle hydrogen-burning at the base of the
outer convective envelope which penetrates inward in regions where He-burning
previously occurred during the interpulse phases in thermally pulsing
intermediate mass stars (hot bottomed convective envelopes: Truran \& Cameron
1971; Pagel \& Edmunds 1981). In this case N-synthesis has a primary-like
behaviour. Lack of an adequate knowledge of the relevant mechanisms prevents
accurate predictions of the run of the [N/Fe] ratio with metal abundance.

The production of Mg (and other $\alpha$-elements like Si, S, Ca, and Ti)
closely mimics the production of O, replacing hydrostatic He-burning with
hydrostatic burning of C (or O, etc.). Core collapse SN explosions of massive
stars are expected to return to the ISM large amounts of Mg (Thielemann et al.
1990; Tsujimoto et al. 1995; Woosley \& Weaver 1995).

Finally, Na production during C-burning in massive stars is primary, it is a
rather strong function of the neutron density (Truran \& Arnett 1971; Woosley
\& Weaver 1995); since this is expected to be a function of initial stellar
metallicity, the [Na/Mg] ratio is predicted to fall below solar in metal-poor
stars (Arnett 1973; Truran 1973; Woosley \& Weaver 1995). It should however be
noted that Na synthesis by p-capture on $^{22}$Ne in regions of H-burning
shell where O is converted into N has been claimed to explain the Na excesses
observed in supergiants and in bright giants in globular clusters (Denisenkov
\& Denisenkova 1990); Woosley \& Weaver (1995) predicts that some 10\% of the
$^{23}$Na synthesized in massive stars is produced in the H-envelope as
consequence of the Ne-Na cycle. The effects of this reaction during the RGB
phase of small mass stars have been studied in detail by Langer et al. (1993).
These authors found that also a $^{20}$Ne-$^{23}$Na cycle might perhaps occur
in these stars; given the large abundances of this last Ne isotope, this
reaction might well explain observed Na excesses. It is now clear that the
Na-O anticorrelation is confined to the large density environments of globular
clusters (see e.g. Gratton et al. 1999b). However no study of the impact of
the possible production of Na in small mass stars on galactic chemical
evolution is available.

Observational evidences concerning these elements are discussed in a number of
reviews (see e.g. Wheeler et al. 1989). In the following, we will only recall
the most important points and give some updates. The situation for C and N is
rather confusing due to difficulties in observing the few good
abundance indices in metal-poor stars, and to the occurrence of mixing in
evolved stars. A variety of spectral indices can be considered for C and N:
however, most of them cannot be used over a wide range of metal abundances.
This is the case of C$_2$\ and CN lines, whose strengths depend approximately
quadratically on metal abundance (neglecting the variation of the gaseous
pressure with metallicity): their observation is limited to evolved and/or
metal rich stars. Several permitted lines of both C and N are observable in
the optical and near-IR; however, they are of high excitation and the possible
presence of departures from LTE should be carefully considered for the
strongest features (see e.g. Tomkin et al. 1992). The forbidden [CI] line at
8727~\AA\ (a high quality abundance indicator in the Sun) is vanishingly weak
in metal-poor stars. Good data are provided by CH; unluckily, the best bands
of NH are in the near UV, in a quite difficult spectral region. Furthermore,
basic data about dissociation and pre-dissociation are not accurately
determined for these molecules (see Lambert 1978), causing offsets to be
present among various abundance determinations. With these caveats in mind, it
is not surprising that there are rather large uncertainties in presently
determined trends with metal abundance: current data suggests that the
abundances of both C (Peterson \& Sneden 1978; Clegg et al. 1981; Laird 1985)
and N (Clegg et al. 1981; Tomkin \& Lambert 1984; Laird 1985; Carbon et al.
1987) scale as the Fe one, with a possible upward trend for C in the most
metal-poor stars (Tomkin et al. 1986; Carbon et al. 1987; Tomkin et al. 1992).
The presence of a moderate trend for decreasing [C/Fe] ratios with increasing
[Fe/H] amongst disk stars ([Fe/H]$>-1$) has been recently proposed by
Andersson and Edvardsson (1994) and Tomkin et al. (1995) from observations of
forbidden and permitted C lines respectively.

The run of O abundances with overall metallicity has been the topic of a rather
large number of papers in the recent past. A nearly constant [O/Fe] ratio at
[O/Fe]$\sim +0.4$\ has been indicated by studies of the forbidden line at
6300~\AA\ in metal-poor giants (Gratton \& Ortolani 1986; Barbuy 1988; Barbuy
\& Erdelyi-Mendes 1989; Sneden et al. 1991a; Kraft et al. 1992). A nearly
constant offset of the abundances of O could be interpreted as the result of
the contribution by type~II SNe alone to the early galactic nucleosynthesis
(see e.g. Matteucci \& Greggio 1986). Larger O excesses have been obtained
from the analysis of the near IR permitted lines in metal-poor dwarfs (Sneden
et al. 1979), with an even larger value ([O/Fe]$\sim 0.9$) obtained by Abia \&
Rebolo (1989). This last analysis does not indicate any plateau in the run of
[O/Fe] with [Fe/H], suggesting that only the ejecta of the most massive among
type~II SNe polluted the medium from which early stars formed, and requiring
then a very fast raise of the metal abundance. Additional important results
have been obtained exploiting OH lines at the UV edge of ground-based and in
the near IR spectra of subdwarfs; note that use of OH bands is very promising,
since these are detectable also in extremely metal-poor stars, where the other
diagnostics are vanishingly weak. Early results (Bessell et al. 1991; 
Balachandran \& Carney 1996) seemed to confirm those obtained from forbiden
lines. However, more extensive analysis (Nissen et al. 1994; Israelian et al.
1998; Boesgaard et al. 1999) are in better agreement with results from the
permitted lines. If these large O abundances held for globular clusters too,
the deduced ages would be substantially reduced.

An examination of these O abundance analyses reveals several weak points. (i)
A Na-O anticorrelation exists amongst the brightest globular cluster giants
(see Kraft 1994 and references therein), although its explanation is still not
clear. In a very recent analysis of mixing episodes amongst field stars
(Gratton et al. 1999b), we find that deep mixing (affecting C and N
abundances) indeed occurs at luminosities brighter than the RGB bump, when the
molecular weight barrier is canceled by the outward expansion of the H-burning
shell. However, O (and Na) abundances are unaltered in field stars, in good
agreement with predictions of theoretical models (Sweigart \& Mengel 1979;
Charbonnel 1994), and previous observational indications (Kraft et al. 1982,
Shetrone 1996, Pilachowski et al. 1996, Kraft et al. 1997). Hence the Na-O
anticorrelation seems to be limited to the dense environments of globular
clusters, and not relevant for the discussion of field stars. (ii) O
abundances are sensitive to the adopted atmospheric parameters. (iii)
Permitted IR lines likely form in non-LTE conditions when lines are rather
strong (Baschek et al. 1977), while no departure from LTE is expected for the
forbidden line at 6300.3~\AA. A very critical test would thus be the analysis
of forbidden lines in very metal-poor dwarfs; unfortunately, this test can be
done only for a few cool, not very metal-poor dwarfs and subgiants: results
(Spiesmann \& Wallerstein 1991; Spite \& Spite 1991; Balachandran \& Carney
1996; Fulbright \& Kraft 1999) support the abundances obtained for giants.
(iv) The UV OH bands used by Israelian et al. (1998) and Boesgaard et al.
(1999) are in a very crowded region in the extreme ground-based UV, and
oscillator strengths for these bands are not well determined: hence, results
from OH cannot be still considered as conclusive (see also Balachandran \&
Bell 1997).

The problem of O abundances in metal-poor stars has been addressed in a series
of papers by King (1993, 1994). He showed that while departures from LTE
likely play only a minor r\^ole (a result also suggested by the statistical
equilibrium computations by Kiselman 1991, 1993), abundances from permitted
and forbidden lines both yield the same [O/Fe] ratio ([O/Fe]$\sim +0.5$, with
only a shallow trend with overall metal abundance) if a new set of effective
temperatures is adopted for subdwarfs and only high quality equivalent widths
($EW$s) are used. Unfortunately, the temperature scale for subdwarfs is still
not well settled: temperatures derived by at least one recent paper (Nissen et
al. 1994) are much lower than those adopted by King (1993); adoption of these
low temperatures would regenerate the very high O abundances found in older
analyses of subdwarfs (see Cavallo et al. 1997).

Abundances for Na and Mg in metal-poor stars have been determined by a number
of authors (Fran\c cois 1986a, 1986b; Magain 1987; Gratton \& Sneden 1987,
1988). However fine details, like the possible existence of a gradient of
[Mg/Fe] with [Fe/H] among stars with [Fe/H]$<-1$, or the precise form of the
run of the [Na/Mg] ratio with [Fe/H], were still not defined in these older
papers at the level required for a detailed comparison with the predictions by
models of the chemical evolution of the Galaxy (se e.g. Matteucci and Fran\c
cois 1992). In part, this is due to the rather large uncertainties present in
the atmospheric parameters used by these otherwise careful discussions. For
Mg, the situation has now changed thanks to the excellent papers by Fuhrmann
et al. (1995), Fuhrmann (1998), and Nissen and Schuster (1997): these
papers showed that (i) Mg is uniformly overabundant in most halo and all thick
disk stars, although there are a few halo stars having lower Mg excesses; and
(ii) the Fe abundance increase at constant Mg at the transition between thick
and thin disk. This last result agrees with our earlier finding for O (Gratton
et al. 1996b), and will be discussed more in detail in Paper IV.

The aim of the present series of papers is to contribute to the knowledge
about the abundances of Fe, C, N, O, Na and Mg in metal-poor stars. In the
course of this investigations, various critical points in the analysis were
examined: some of them are discussed here, while others (the solar abundances
and the applicability of Kurucz model atmospheres in abundance analyses) were
considered in a parallel study of the spectra of RR Lyrae at minimum light
(Clementini et al. 1995) and in a rather extensive comparison with solar
observations (Castelli et al. 1997), in Gratton et al. (1996a: Paper I) (the
\teff\ scale and the adopted atmospheric parameters), and in Gratton et al.
(1999a: Paper II) (non-LTE effects). In this paper we present the abundance
analysis and its results.  Detailed comparisons with nucleosynthesis
predictions and models of galactic chemical evolution models will be done in
forthcoming papers (Gratton et al. 2000: Paper IV; Carretta \& Gratton 2000:
Paper V). The observational material is described in Section~2, where we show
its high degree of internal consistency, with the $EW$s generally accurate to
within $\pm 2$~m\AA. The abundance indicators used in the analysis, with the
basic atomic and molecular data adopted, are described in Section~3. Our final
abundances are presented in Section~4; for O, Mg and Na they include non-LTE
corrections determined in Paper II. We found that our results have a high
internal consistency, at least for \teff$>4600$~K, with error bars in most
cases below 0.1 dex. In particular, we obtain good agreement between abundances
provided by low excitation, forbidden and high excitation, permitted O~I lines;
between C abundances deduced from C$_2$, CH, permitted and forbidden C~I lines;
and between Na abundances provided by the D and subordinate lines. Some problem
still exists for Mg. We compare our results with those from some recent paper
in Section 5; finally  Conclusions are given in Section~6. Finally, note that
the material on which the present analysis is based on does not include the
data discussed in Gratton et al. (1999b).

\section{Observational material } 

The observational material is described in paper I; here we recall that the
original data concerning 19 metal-poor stars were complemented by literature
data in order to better examine trends of abundances with [Fe/H] and
luminosity. We considered selected data sets giving $EW$s for both neutral and
ionized Fe lines, in order to derive gravities from ionization equilibrium for
all stars. The following data sets were considered:
\begin{itemize}
\item Tomkin et al (1992, hereinafter TLLS) analysis of high excitation,
permitted C and O lines in 34 unvolved metal-poor stars. \item Sneden et al.
(1991) and Kraft et al. (1992: hereinafter collectively SKPL) analysis of
forbidden O and Na lines of field and cluster halo giants; for this data set
we only considered here field stars (27 objects), owing to possible systematic
differences between field and cluster stars.
\item Edvardsson et al. (1993; hereinafter E93) analysis of permitted high
excitation O, Na and Mg lines in 187 stars with [Fe/H]$>-1$. Data by E93
were complemented by those of Clegg et al. (1981), Andersson and Edvardsson
(1994) and Tomkin et al (1995) for C lines, and Nissen and Edvardsson (1992)
for the [O~I] forbidden lines in a smaller number of stars. 
\item Zhao \& Magain (1990, hereinafter ZM90) data for Na and Mg in 20
metal-poor dwarfs. 
\end{itemize}

Atmospheric parameters for the program stars are given in Table~2 of Paper~I.
While data were obtained with different instrumentation and reduction
procedures, the final set of $EW$s is highly homogenous. Mean differences
between $EW$s determined on our original sample and those from literature are:
$$EW_{\rm us}-EW_{\rm E93}=2.5\pm 0.5~{\rm m\AA}~(\sigma=2.8~{\rm m\AA)},$$ 
$$EW_{\rm us}-EW_{\rm SKPL}=2.4\pm 0.6~{\rm m\AA}~(\sigma=3.5~{\rm m\AA)},$$
$$EW_{\rm us}-EW_{ ZM90}=-0.4\pm 0.7~{\rm m\AA}~(\sigma=3.3~{\rm m\AA)},$$
based on 21, 34 and 27 lines respectively. We did not apply any systematic
corrections to the original $EW$s, and we estimate typical errors in individual
$EW$s to be $\pm 2$~m\AA. 

\section{Abundance indicators and analysis} 

Abundances were derived using model atmospheres extracted from the grid by
Kurucz (1992; hereinafter K92)\footnote{ CD-ROM 13; these models used here have
the overshooting option switched on} and compared with solar
ones obtained using the solar model atmosphere from the same grid, with a depth
independent microturbulent velocity of \vt=0.9~\kms (Simmons \& Blackwell
1982). These solar abundances are very close to those obtained using the
Holweger \& M\"uller (1974) empirical solar model atmosphere (which is
currently the best representation of solar photosphere), and to the meteoritic
values listed by Anders \& Grevesse (1989). Fe abundances are listed and
discussed in Paper I; here, [Fe/H] values will be repeated only when needed. 

CNO abundances were obtained using molecular concentrations given by a
simultaneous solution of the dissociation equations for several species,
following the method described in Lambert \& Ries (1977, 1981). Coupling of
atomic C and O concentrations due to CO formation is of particular relevance
for C. We used O abundances given by the average of permitted and forbidden
lines (see below). Furthermore, CN line strength depends on the product of the
concentration of atomic C and N: hence, any change in the C abundance causes a
corresponding modification in deduced N abundances. Corrections for CO
formation are less important for O (since n(O)$>>$n(C) in all program stars),
the difference between abundances computed with [C/Fe]=0 or [C/Fe]$=-0.5$\
being less than 0.05~dex in most cases. Given these interrelations among the
various species, we will first present our results for O, and then discuss
those for C and N. 

\begin{table*}
\caption[ ]{Effects of errors in the atmospheric parameters for a metal-rich 
dwarf and a metal-poor giant }
\label{tab:1}
\begin{flushleft}
\begin{tabular}{llrrrcrc}
\hline\noalign{\smallskip}
\multicolumn{1}{c}{Element} 
&\multicolumn{1}{c}{Index} 
&\multicolumn{1}{c}{${\rm \Delta ~ T_{eff}}$}  
&\multicolumn{1}{c}{$\Delta ~\log~ g$}
&\multicolumn{1}{c}{${\rm \Delta[A/H]}$}
&\multicolumn{1}{c}{$\Delta~ v_t$}  
&\multicolumn{1}{c}{$\Delta_{tot}$}
&\multicolumn{1}{c}{$\Delta$(Sun)} \\
	 &       &+100K  &$-$0.5 dex & $-$0.2 dex & +0.5 km/s &&\\
&&&&&&\\
\noalign{\smallskip}
\hline\noalign{\smallskip}
&&&&&&\\
    HD 102365  &&&&&&\\
\\
${\rm [Fe/H]}$  & Fe I   &     0.10&$-$0.03&$-$0.04&$-$0.04& 0.12&   ~0.16\\
${\rm [Fe/H]}$  & Fe II  &  $-$0.02&$-$0.30&$-$0.09&$-$0.04& 0.32&   ~0.04\\
${\rm [O/Fe]}$&${\rm [OI]}$&   0.03&   0.02&   0.09&  ~0.04& 0.10& $-$0.05\\
${\rm [O/Fe]}$  & OI     &  $-$0.07&   0.11&   0.11&  ~0.03& 0.17& $-$0.10\\
${\rm [C/Fe]}$&${\rm [CI]}$&   0.01&   0.09&   0.02&  ~0.04& 0.10& $-$0.02\\
${\rm [C/Fe]}$  & CI     &  $-$0.04&   0.12&   0.09&  ~0.04& 0.16& $-$0.10\\
${\rm [C/Fe]}$  & CH     &  $-$0.05&$-$0.01&   0.02&  ~0.04& 0.07&   ~0.00\\
${\rm [C/Fe]}$  & C2     &  $-$0.11&$-$0.09&   0.04&  ~0.04& 0.15& $-$0.11\\
${\rm [N/Fe]}$  & CN     &  $-$0.05&$-$0.06&   0.05&  ~0.04& 0.10& $-$0.22\\
${\rm [Na/Fe]}$ & Na I   &  $-$0.04&   0.06&   0.00&  ~0.03& 0.08& $-$0.04\\
${\rm [Mg/Fe]}$ & 4571   &     0.04&   0.04&   0.00&$-$0.01& 0.06& $-$0.01\\
${\rm [Mg/Fe]}$ & others &  $-$0.05&   0.02&   0.01&  ~0.04& 0.07& $-$0.06\\
\\
    HD 122956 &&&&&&\\
\\
${\rm [Fe/H]}$  & Fe I   &    0.17 &    0.06 &    0.02 & $-$0.21 & 0.21&\\
${\rm [Fe/H]}$  & Fe II  & $-$0.03 & $-$0.17 & $-$0.04 & $-$0.11 & 0.18&\\
${\rm [O/Fe]}$&${\rm [OI]}$&  0.03 & $-$0.04 &    0.06 &   ~0.10 & 0.09&\\
${\rm [O/Fe]}$  & OI     & $-$0.10 & $-$0.03 &    0.08 &   ~0.08 & 0.14&\\
${\rm [C/Fe]}$  & CH     & $-$0.13 & $-$0.04 &    0.01 &   ~0.18 & 0.16&\\
${\rm [C/Fe]}$  & C2     & $-$0.17 & $-$0.16 & $-$0.01 &   ~0.21 & 0.26&\\
${\rm [N/Fe]}$  & CN     & $-$0.12 & $-$0.19 & $-$0.02 &   ~0.24 & 0.26&\\
${\rm [Na/Fe]}$ & Na I   & $-$0.10 & $-$0.02 & $-$0.01 &   ~0.20 & 0.14&\\
${\rm [Mg/Fe]}$ & 4571   &    0.08 &    0.02 &    0.03 & $-$0.10 & 0.10&\\
${\rm [Mg/Fe]}$ & others & $-$0.11 & $-$0.03 & $-$0.01 &   ~0.20 & 0.15&\\
\\
\noalign{\smallskip}
\hline
\end{tabular}
\end{flushleft}
\end{table*}

Table~\ref{tab:1} presents an analysis of the uncertainties on the derived
abundances related to possible errors in the adopted atmospheric parameters.
This was obtained by repeating the whole procedure of abundance derivations by
modifying one parameter at a time for a couple of typical cases (a metal-rich
dwarf and a metal-poor giant). Oxygen abundances were compared with those
given by Fe~II lines, in order to reduce their sensitivity on the
adopted gravities. Column~7 of Table~\ref{tab:1} lists a total uncertainty
(due to possible errors in the atmospheric parameters); this was given by a
quadratic sum of the effects of errors in the individual parameters\footnote{
An error of $\pm 0.25$~\kms\ in \vt\ was considered for giants, since for these
stars \vt\ is very well constrained from our spectra}. Column~8 shows how the
abundances change when solar abundances obtained using the HM model atmosphere
are replaced by those given by the solar model by Bell et al (1976). Values
listed in this column provide a guess about uncertainties related to the
structure of the model atmosphere.

Comparison of the values listed in Table~\ref{tab:1} with those given in the
analogous tables of Gratton \& Sneden (1991, 1994) (where typical overall
uncertainties of the element-to-iron abundance ratios were 0.05 -- 0.08~dex)
reveals that the abundances for the elements considered in this paper are often
much more sensitive on uncertainties in the atmospheric parameters (due to
different sensitivities on temperature and pressure) and in the structure of
the model atmosphere (since lines form at various depths). In particular, the
concentration of various molecules (like C$_2$\ and CN) is very sensitive to
temperature and pressure in the outer (and cooler) regions of the stellar
atmospheres. On the other hand, O and C abundances deduced from high
excitation, permitted lines are strongly sensitive to temperature in the
deepest part of the atmospheres, and hence on how convection is handled when
computing models. Finally, the abundance ratios are sensitive to the adopted
values for the microturbulent velocities because thermal motions are larger
than microturbulent velocities for these light species. The smaller overall
sensitivity to uncertainties in the atmospheric parameters and model
atmospheres of O and C abundances deduced from [OI] and CH lines support their
use as primary abundance indicators. 

\begin{table}
\caption[ ]{ Equivalent widths for O lines and Na lines in the 19 stars of our
sample. Available only in electronic form }
\label{tab:2}
\end{table}

\subsection{Oxygen}

The present oxygen abundances are based on the forbidden line at 6300.3~\AA,
which is considered a first class abundance indicator (Lambert 1978), and on
the analysis of the high excitation, permitted triplets at 615.8 and 777~nm,
including a complete treatment of the non-LTE effects on the abundances derived
from this last two indicators (see Paper II for an exhaustive discussion about
these statistical equilibrium calculations). However, since there is still no
general agreement about non-LTE corrections, we listed also the abundances
obtained from the LTE analysis. Oscillator strengths $gf$'s
were from Garstang (1976) for the forbidden line, and from Bi\'emont et al
(1991) for the permitted ones. $EW$s for O lines in the 19 stars of our sample
are listed in Table~\ref{tab:2} (only available in electronic form). 

O abundance determinations made (as in the present paper) using both permitted
and forbidden lines ought to be more reliable than those using just one of
these transitions. [O~I] lines are weak in warm subdwarfs; on the other side,
lines of the high excitation triplet are weak in cool giants. Analyses based
on forbidden lines in metal-poor giants and on the IR triplets in subdwarfs
probably introduce significant biases (see also Gratton 1990; 1993).

\subsection{Carbon}

Carbon abundances for the stars in our original data set were derived from a 
comparison of observed and synthetic spectra of the wavelength region 
4207--4225~\AA, which includes several features due to the R branch of 
the $A^2\Pi -X^2 \Sigma$\ ($\Delta v=0$) transition of CH. For a few, 
generally metal-rich stars, additional information were provided by a similar 
comparison for the spectral feature at 5086~\AA, mainly due to lines of the 
C$_2$\ Swan system. Unfortunately, C$_2$\ lines are vanishingly weak in most 
metal-poor stars. Both these sets of abundance indices were considered by
Lambert (1978) in his analysis of the solar Carbon abundance: Lambert derived
values of log~n(C)=8.67 and 8.73 from the CH $A-X$\ and the C$_2$\ Swan bands
respectively. While Lambert observed that the CH $A-X$\ band cannot be
considered the best abundance indicator in the Sun (due to concern in the
correct location of the continuum level at these short wavelengths), it is the
best choice in metal-poor stars, as shown by Sneden et al. (1986) and TLLS. In
particular, TLLS showed that a systematic error is possibly present in their C
abundances obtained from C~I lines around 9100~\AA: the derived abundances show
a residual trend with T$_{\rm eff}$ that vanishes when the C abundance is
instead obtained from CH (see their Fig. 8) \footnote{ We will later show that
this trend is canceled when our \teff\ scale is adopted}. 

The dissociation potential for CH has been determined with high accuracy at
$D_o^o=3.464$~eV (Brzozowski et al. 1976). Unfortunately, situation is not
so good for the band oscillator strengths (see discussions in Grevesse \&
Sauval 1973 and Chmielewski 1984). Furthermore, predissociation greatly
affects the strength of transitions with rotational quantum numbers $R$\ larger
than $R\sim 20$\ for the (0,0) band, $R\sim 10$ for the (1,1) band, and with
any value of $R$\ for the (2,2) band. Hence, we prefer to use the band
oscillator strengths deduced by Chmielewski (1984) from an analysis of the
solar spectrum. The lines considered by Chmielewski have rotational quantum
numbers similar to those considered here (actually the set of lines used in his
study overlaps with our), and therefore use of his recommended values is
appropriate here. Since the band oscillator strengths are taken from a solar
analysis, it is not surprising that the synthetic spectra show good
consistency with the KPNO solar flux spectrum (Kurucz et al. 1984). 

The basic parameters for the synthesis of the C$_2$\ feature at 5086~\AA\
were taken from Lambert \& Ries (1981). Again, a comparison with the solar flux
spectrum yields a good agreement. 

In the reanalysis of literature data, we considered only $EW$s for atomic lines,
since spectral profiles required for comparisons with synthesis of molecular
features were not available. Three groups of features were considered:
permitted high excitation lines close to 9100 (Clegg et al 1981; TLLS) and
7100~\AA\ (Tomkin et al 1995), and the forbidden line at 8727~\AA\ (Andersson
and Edvardsson 1994). Features of the first group are stronger, and can be 
observed in very metal poor stars, while the others are weak and can only be
observed in disk stars. In the analysis of these lines we used line data from
the original papers: it should be noticed that the solar abundances obtained
with the permitted lines near 7100~\AA (log~n(C)=8.66) and the forbidden line
(log~n(C)=8.67) agree very well with the recommended solar abundance of
log~n(C)=8.60 (Grevesse et al 1991). 

While the effects of departures from LTE are certainly negligible for the
weak [CI] line, they might be more significant for the high excitation
permitted line. Statistical equilibrium calculations for the C features
considered in this paper have been carried out by St\"urenburg \& Holweger
(1990) and Tomkin et al (1992, 1995). These calculations indicate that
non-LTE abundance corrections are small ($<0.1$~dex) for the lines 
near 7100~\AA\ (hence we simply adopted the LTE values). However, they
may be not negligible for the stronger lines near 9100~\AA, depending on the
adopted cross sections for collisions with HI atoms (TLLS). To take into
account these non-LTE effects, we corrected the average C abundances from the
reanalysis of TLLS $EW$s by the same amount they did. 

\subsection{Nitrogen}

We obtained nitrogen abundances using a number of CN features due to the
$\Delta v=-1$\ bands of the violet system ($B^2\Sigma -X^2\Sigma $) near the
bandhead at 4215~\AA\ \footnote{The bandhead itself is severely blended with
the strong resonance line of Sr~II and cannot be used}, and the 2,0 vibrational
band of the red system ($A^2\Pi -X^2\Sigma$) in the wavelength range 7910-7982~
\AA. Unluckily, no observation were available for the NH bands. The
dissociation potential of CN is not well determined (see e.g. the discussions
in Lambert 1978, and Lambert \& Ries 1981), with values ranging from 7.5 to
7.9~eV: this is still the major source of uncertainty in the derived abundances
of N from CN. In fact, the above mentioned range translates in a factor of
about 3 in the derived N abundances. Here we will assume a value of
$D_o^o=7.65$~eV (Engleman \& Rouse 1975). The $f_{00}$\ oscillator strength of
the violet band was taken from Danylewych \& Nicholls (1978), while the
Franck-Condon factors were from Dwivedi et al. (1978). Band strengths of the
red system are discussed in Sneden \& Lambert (1982), who also compared
theoretical values with those derived from analysis of the solar spectrum. Here
we adopted their solar oscillator strengths for the (2,0) band, and verified
that they well reproduce the KPNO solar flux spectrum.

\subsection{\C12C13 ratio}

The \C12C13 ratio was determined by a spectral synthesis of three $^{13}$CH
features at 4211.46, 4213.13, and 4221.83~\AA, since most of the higher quality
$^{13}$CH indicators listed by Sneden et al. (1986) were not observed, and
$^{12}$C$^{13}$C and $^{13}$CN lines are extremely weak in the spectra of the
program stars. The positions of $^{13}$CH lines were computed using the same
code and molecular parameters considered by Sneden et al., while concentration
was obtained using the same precepts adopted for $^{12}$CH. Comparison of
synthesized spectra with the solar one gives a good match for the canonical
solar \C12C13 ratio of 89; however, this comparison is not so meaningful here,
due to the large solar \C12C13 ratio. We may however compare the present
\C12C13 ratios with those determined by Sneden et al for 3 giants in common
between the two samples: on average, our \C12C13 ratios are larger by $24\pm
14$\% ($\sigma=$25\%). We attribute this difference to our use of not first
class indicators; however, we will adopt our \C12C13 ratios with no further
correction. 

\subsection{Sodium }

Na abundances for stars in the original sample mainly rest on the $EW$s for the
yellow doublet at 5682 and 5688~\AA, for which we performed a full non-LTE
analysis (see Paper II for details). However, since there is still no
general agreement about non-LTE corrections, we listed also the abundances
obtained from the LTE analysis. Additional information is provided by the
line at 6154~\AA\ (the other component of this doublet has not been observed),
and by the weak line at 4751~\AA\ (see Table~\ref{tab:2}). All these lines are
listed as clean lines in the Holweger (1971) analysis of the photospheric Na
abundance; the profile of the 5682.647~\AA\ line shows actually some
distortion in the short wavelength side in the solar spectrum, due to
contamination by a weak, high excitation (E.P.=3.84~eV) Cr~I line at
5682.493~\AA. However, synthetic spectra for both the Sun and the program
stars show that this contamination does not affect significatively the abundance
derivation: the mean difference between abundances extracted from the 5682 and
the (clean) 5688~ \AA\ line is only $0.04\pm 0.02$\ dex ($\sigma =0.10$~dex).
In Table~\ref{tab:2} we listed the $EWs$ measured for Na in the program stars.
Data gathered from the literature include $EW$s for both the 5682-88~\AA\ and
6154-60~\AA\ doublets.

$gf$s for all sodium lines can be accurately computed (see e.g. Lambert \&
Warner 1968). Of more concern is the uncertain value of collisional damping
parameters, since lines used in the present analysis are rather strong in
metal-rich stars. Holweger (1971) obtained a rather large value of the
enhancement factor for Na lines ($\log E=0.8$) by ruling out any dependence of
the abundances on line strength; we obtained a very similar value ($\log
E=0.73\pm 0.14$) by repeating the same procedure with our code, the HM solar
model atmosphere, and a depth independent microturbulent velocity of 0.9~\kms\
(Simmons \& Blackwell 1982). The solar sodium abundance provided by the program
lines is log~n(Na)=6.31, identical to the meteoritic value (Anders \& Grevesse
1989); however, for consistency with a forthcoming study of globular cluster
giants (Carretta \& Gratton 2000), we adopted a solar value of log~
n(Na)=6.33, as derived from Anders \& Grevesse (1989).

\subsection{Magnesium } 

Our magnesium abundances are based on the red triplet at 6318--6319~\AA, on
the singlet line at 4730~\AA, and on the intercombination line at 4571~
\AA. This last line is very clean in the solar spectrum, and it has been used
by various authors as a diagnostic of the solar chromosphere (see e.g. Mauas et
al. 1988); there has been however some confusion in the astronomical literature
about its $gf$: e.g. Zhao \& Magain (private communication) used a value of
$\log gf=-5.44$, which yields a discrepant value for the solar Mg abundance.
However, the $gf-$value for this line of $\log gf=-5.69$ has been determined
from accurate lifetime measurements (Kwong et al 1982): a solar analysis with
this value of the $gf$, an $EW$\ of 101~m\AA\ (Steffen 1985) and the K92 solar
atmosphere yielded a value of the abundance of Mg of log~n(Mg)=7.65,
in good agreement with that obtained from the other lines in our
list (log~n(Mg)=$7.60\pm 0.02$) and with the meteoritic Mg abundance of
log~n(Mg)=$7.58\pm 0.02$ (Anders \& Grevesse 1989). The other lines employed
in our analysis are listed as clean by Lambert \& Luck (1978), and have
accurate $gf$'s (Froese-Fisher 1975a, 1975b; Mendoza \& Zeippen 1987). However
the red triplet is so close that it is not well resolved even on our rather
high resolution spectra; and the 4730.038~\AA\ line is blended with a Cr~
I line at 4729.859~\AA, so that accurate $EWs$ are not easily derived
for these features. We then adopted the abundances given by comparison with
synthetic spectra. Reanalysis of literature data uses $EW$s both for the
intercombination and high excitation Mg~I lines. 

\begin{table}
\caption[ ]{CNO abundances and \C12C13 ratios for the 19 program stars
(available only in electronic form) }
\label{tab:3}
\end{table}

\begin{table}
\caption[ ]{Reanalysis of data by Tomkin et al. (1992): TLLS 
(available only in electronic form)}
\label{tab:4}
\end{table}

\begin{table}
\caption[ ]{Reanalysis of data by Sneden et al. (1991) and Kraft et al (1992): 
SKPL(available only in electronic form) }
\label{tab:5}
\end{table}

\begin{table}
\caption[ ]{ Reanalysis of data by Edvardsson et al. (1993): E93 
(available only in electronic form)}
\label{tab:6}
\end{table}

\begin{table}
\caption[ ]{Reanalysis of data by Zhao \& Magain (1990): ZM90
(available only in electronic form) }
\label{tab:7}
\end{table}

\section{Results }

CNO abundances and \C12C13 for our program stars are listed in
Table~\ref{tab:3}; Tables from \ref{tab:4} to \ref{tab:7} list the results from
the reanalysis of literature data (TLLS, SKPL, E93 and ZM90, respectively). 
These tables are available only in electronic form.

\subsection{Oxygen }

All our results for O include corrections for non-LTE effects obtained
following the procedure described in Paper II, $i.e.$ including statistical
equilibrium calculations based on empirically calibrated collisional H~I cross
sections. 

\begin{figure}
\psfig{figure=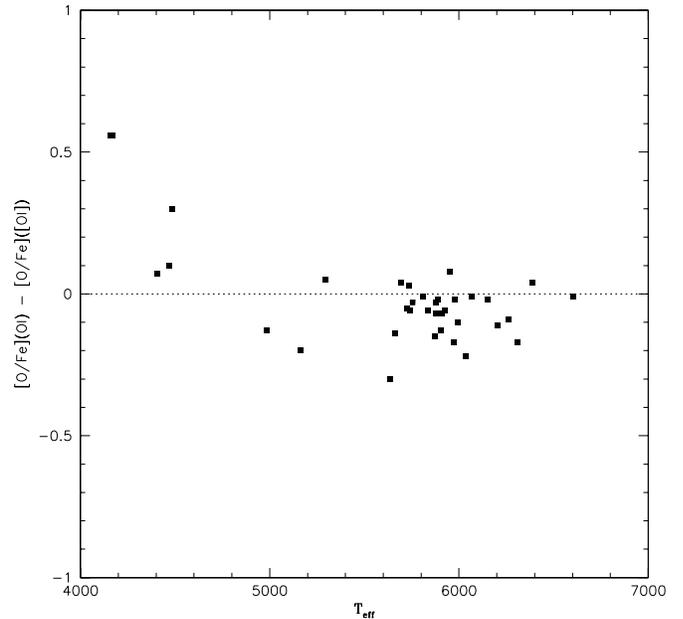,width=9.0cm,clip=}
\caption[ ]{Run of the [O/Fe] difference from permitted and forbidden lines 
against T$_{\rm eff}$} 
\label{fig:2}
\end{figure}

Mean residual between O abundances deduced from permitted and forbidden
lines is $0.01\pm 0.03$~dex ($\sigma=0.18$~dex; 40 stars). This difference
is plotted against \teff\ in Figure~\ref{fig:2}. 

As one can see, residuals for the coolest stars  are much larger than average;
this can be attributed to inadequacy of K92 models with the overshooting
option switched on to describe the atmospheres of these cool stars (see Paper
I and discussion below). Note that better agreement between different O
diagnostics as well as a better equilibrium of ionization for Fe are obtained
using the Kurucz models with no overshooting used in Gratton et al. (1999b);
however, since these models became available to us only after all computations
for the present paper were completed, and since abundances relative to the
Sun obtained with the two sets of models agree very well each other for stars
warmer than \teff=4600~K, we decided not to repeat these lengthy computations
and simply eliminated stars cooler than this limit from further discussion.
Once this was done, the mean residual between O abundances provided by
permitted and forbidden lines is $-0.06\pm 0.02$~dex (32 stars;
$\sigma=0.15$~dex); both mean difference and scatter are very small. Note that
in our analysis we assumed a solar O abundance of log~n(O)=8.93, the value
recommended by Anders \& Grevesse (1989) from an analysis of the forbidden
lines and OH bands. However, a study of permitted lines in the Sun (Bi\'emont
et al. 1991) gives a solar O abundance of log~ n(O)=8.86; the difference with
the abundance from forbidden lines found in the Sun is then almost the same
we found in field stars.

Of course, this result depends on the adopted non-LTE corrections, which stem
from our own statistical equilibrium calculations. However, we found that these
corrections are small: on average, abundances derived from IR permitted lines
must be lowered by 0.06~dex, the largest correction ($\sim 0.25$~ dex) being
obtained for the warmest stars. Explorative computations with $x$=0.03 and
$x$=3 (the extreme values of the multiplicative factor $x$\ for H~I collisions
compatible with observations of RR Lyrae; see Paper II for the definition of
$x$) show that the non-LTE corrections are uncertain by about $\pm 0.1$~dex for
these stars.

\begin{figure}
\psfig{figure=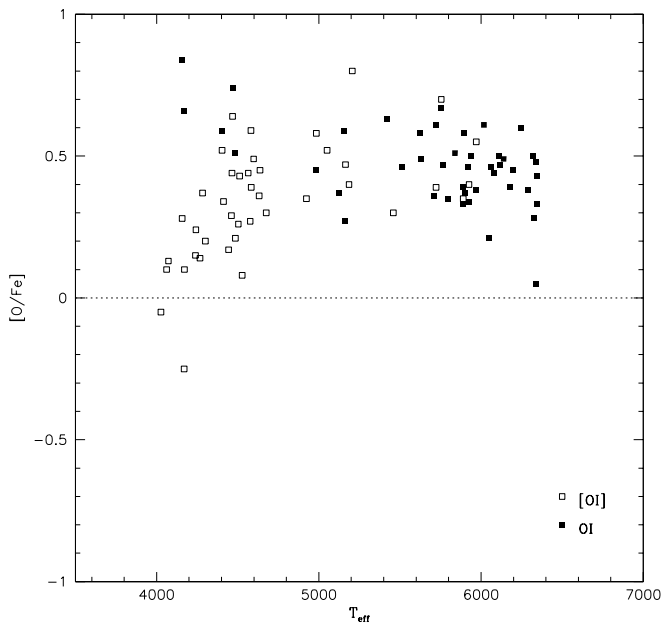,width=9.0cm,clip=}
\caption[ ]{Run of the [O/Fe] ratios with T$_{\rm eff}$ in metal-poor stars 
([Fe/H]$< -0.7$. Different symbols are used for abundances derived from 
permitted 
(filled symbols) and forbidden lines (open symbols)} 
\label{fig:3}
\end{figure}

Figure~\ref{fig:3} shows the run of the [O/Fe] ratio with T$_{\rm eff}$ for
stars with [Fe/H]$< -0.7$, both from permitted and forbidden lines. Systematic
trends in abundances derived from both indicators are clearly present. These
trends cannot be attributed to an evolutionary effect for two reasons: 
\begin{enumerate}
\item Permitted and forbidden lines exhibit opposite trends
\item Three of the stars in the SKPL sample are likely to be in
evolutionary phases later than the RGB, according to their position in the
$(b-y)-c_1$\ diagram of Twarog \& Anthony-Twarog (1994): two of them are
probably horizontal branch stars (BD+11$^0$2998: [O/Fe]=+0.30; and HD~166161:
[O/Fe]=+0.40), while the third one is likely to be at the base of the asymptotic
giant branch (HD~204543: [O/Fe]=+0.36). Their [O/Fe] ratios are much larger
than those measured in several cooler stars, and actually do not differ
significantly from the average obtained for warm stars. 
\end{enumerate}

Departures from LTE are not expected for the forbidden lines. Hence, the most
likely explanation for the trend with $T_{\rm eff}$\ is large deviations of the
atmospheric structure from that of K92 models used in this paper, that seem to
be not fully adequate to reproduce the coolest stars. This last possibility is
also supported by the poor equilibrium of ionization we get for Fe (Paper I).

\begin{figure}
\psfig{figure=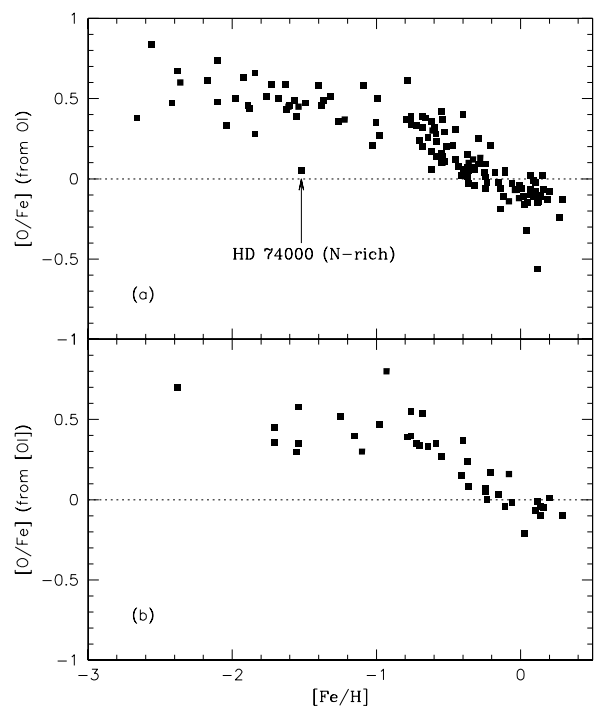,width=10.0cm,clip=}
\caption[ ]{Run of the [O/Fe] ratios vs [Fe/H] in metal-poor field stars. [O/Fe] 
ratios are obtained from permitted (panel a) and from forbidden (panel b) 
lines; only stars with \teff$>4600$~K were considered. The N-rich dwarf 
HD~74000 is indicated, while the other N-rich dwarf present in our samples, 
HD~25329, has been omitted, since it has only an upper limit for O abundance} 
\label{fig:4}
\end{figure}

It seems then safer to consider only warm stars when examining the run of the
[O/Fe] ratio with [Fe/H]. Panels~a and b of Figure~\ref{fig:4} display the run
of the [O/Fe] ratio against overall metal abundance [Fe/H] for permitted and
forbidden lines respectively; only stars with \teff$>4600$~K were considered.
The two runs appear to be very similar; while a small upward trend in [O/Fe]
among the most metal-poor stars may exist (somewhat in agreement with the recent
results based on the OH band by Israelian et al. 1998, and Boesgaard et al.
1999, although less pronounced than found by these authors), we will
hereinafter adopt an average value for the O overabundance in all metal-poor
stars. This average value (from forbidden lines) in warm metal-poor stars is:
$${\rm [O/Fe]}=+0.48\pm 0.05~{\rm dex}~~~(\sigma=0.16~{\rm dex,~11~stars}),$$
which is indeed very close to the value obtained from permitted lines: 
$${\rm [O/Fe]}=+0.45\pm 0.02~{\rm dex}~~~(\sigma=0.13~{\rm dex,~33~stars}),$$
This agreement is due to the adoption of a T$_{\rm eff}$'s scale
about 100--150~K higher than the ones (e.g. Carney 1983; Magain 1987) used in
previous abundance analysis of subdwarfs (see also Paper I). Note that in
no case we obtain [O/Fe]$>1$; hence we cannot confirm the large O
overabundances found by Abia \& Rebolo (1989) for the most metal metal-poor
stars. On the basis of our results and of various discussions in the literature
(e.g. TLLS) we confirm that their overestimates are mainly due
to errors in the $EW$s and to the adoption of Carney (1983) \teff's scale (see
also King 1993). The use of a linear run of [O/Fe] vs [Fe/H], with no $plateau$
at low metallicity should be strictly avoided when interpreting the
colour-magnitude diagrams of globular clusters, in order to obtain reliable
distances and ages (see Buonanno et al. 1989 for a discussion of this point). 

>From our results, we conclude that O abundances derived from permitted and
forbidden lines are in good agreement each other, for stars with \teff$>4600$~K
and in the following we will adopt the average abundances. The mean
overabundance of O in field stars, derived both from permitted and forbidden
lines, is then: 
$${\rm [O/Fe]}=+0.46\pm 0.02~{\rm dex}~~~(\sigma=0.12~{\rm dex,~32~stars}),$$
This mean overabundance also agrees with other recent estimates: [O/Fe]=0.41
from the forbidden lines in the spectra of giants (Sneden et al. 1991); and
[O/Fe]=0.52 from the permitted lines in the spectra of dwarfs (King 1993). 

\subsection{Carbon }

Carbon abundances provided by CH and C$_2$\ lines for our original star list
are listed in Columns~3 and 4 of Table~\ref{tab:3} respectively. Abundances
from the C$_2$\ blend at 5086~\AA\ are available only for 6 metal rich dwarfs;
on average, they are slightly larger than those given by the CH lines in the
range 4203--4224~\AA\ by $0.08\pm 0.02$~dex (the r.m.s. scatter is 0.05~dex).
This systematic difference is small, although it seems to be
significant\footnote{In the analysis of Lambert (1978) of the solar C
abundance, the difference between the values obtained from the same transition
considered here ranges from 0.06 to 0.15~dex, depending on the model
atmosphere used. The HM model gives a difference of 0.09~dex, in agreement
with the value we found within observational errors.}. However, the large
differences between the C abundances obtained using C$_2$\ lines when
different solar model atmospheres are considered (see Table~\ref{tab:1}) might
suggest that the structure of the model atmospheres plays an important r\^ole
here. Luckily, [C/Fe] ratios deduced from CH lines are much less sensitive on
the adopted model atmospheres: hence, they will be considered in the following
discussion.

Carbon abundances from atomic lines for the reanalyzed samples are listed in
Table~\ref{tab:4} (data for the TLLS sample) and ~\ref{tab:6} (those for the 
E93,
obtained using both permitted and forbidden lines: see Clegg et al 1981; Tomkin
et al 1995; and Andersson and Edvardsson 1994). For 12 stars we have data both
from low excitation forbidden and high excitation permitted lines: abundances
from permitted lines are on average larger by $0.03\pm 0.05$~dex
($\sigma=0.19$~dex). The mean difference is very small, but there might be a
small trend for increasing differences with decreasing \teff. A larger sample
is required to explore this issue. 

\begin{figure}
\psfig{figure=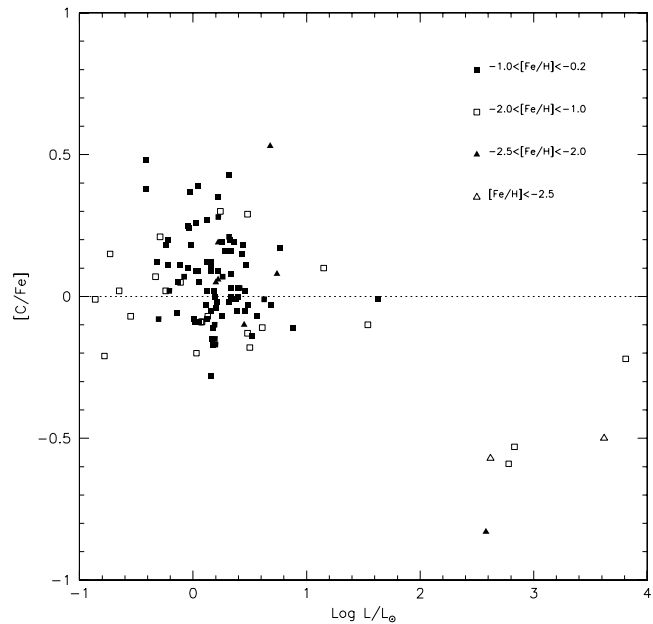,width=9.0cm,clip=}
\caption[ ]{Run of the [C/Fe] ratio against stellar luminosity for the stars
analyzed in the present paper. Different symbols represent stars in
different bins of metal abundance} 
\label{fig:5}
\end{figure}

\begin{figure}
\psfig{figure=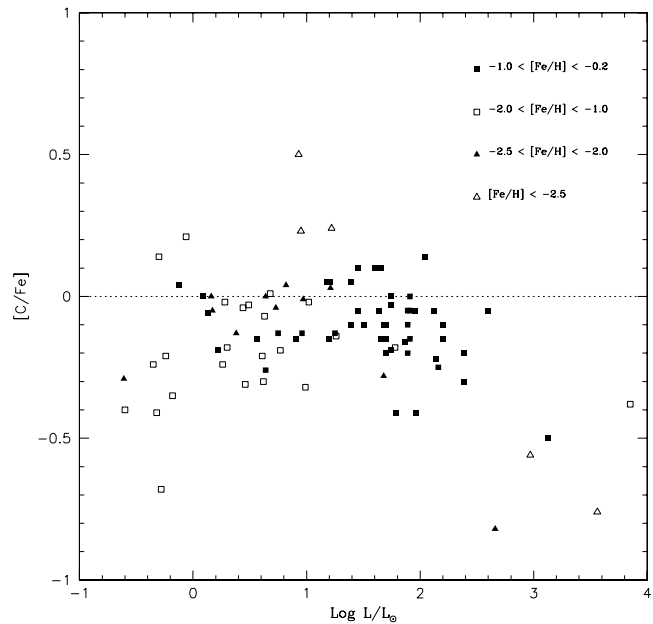,width=9.0cm,clip=}
\caption[ ]{Run of the [C/Fe] ratio against stellar luminosity in metal-poor
stars ([Fe/H]$<-0.2$) from a compilation of selected studies (Cottrell \&
Sneden 1986; Luck 1991; this paper). Different symbols represent stars in
different bins of metal abundance} 
\label{fig:6}
\end{figure}

Before discussing the run of C abundances with overall metallicity, stars which
have undergone mixing episodes must be eliminated, since C is expected to be
depleted in luminous giants (see Figure~\ref{fig:5}, where we used different
symbols for stars in different bins of metal abundances). However, the precise
location of the first dredge up cannot be determined from these data alone,
due the lack of stars of intermediate luminosities ($1<\log L/\Lsol<2.5$).
This issue is considered in detail in another analysis (Gratton et al.
1999b). In Figure~\ref{fig:6} we have plotted [C/Fe] ratios determined from
the G-band in the present work and from a compilation of literature studies.
While the scatter of these [C/Fe] ratios is large (data include C abundances
determined from low dispersion spectra as well as higher quality
determinations), this figure reveals that the depletion of surface C
abundances occurs at a luminosity around $\log L/\Lsol\sim 2$, or slightly
fainter, in agreement with predictions from evolutionary models (Sneden et al.
1986), and with observations of stars in metal-poor globular clusters (see
e.g. Bell et al. 1990), and with the results of Gratton et al. (1999b). Since
there is no evidence that any appreciable C depletion has occurred for stars
as bright as $\log L/\Lsol\sim 1.5$, in the following we will assume that
stars fainter than this limit have the original surface C abundance.

\begin{figure}
\psfig{figure=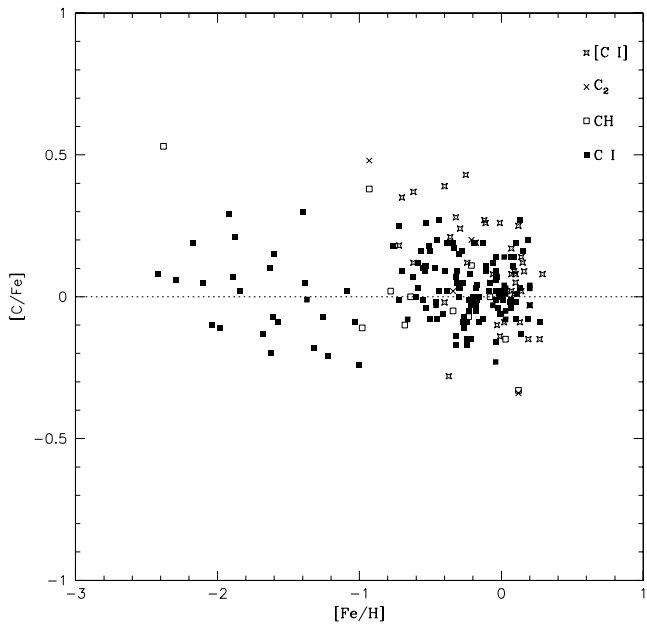,width=9.0cm,clip=}
\caption[ ]{Run of the [C/Fe] ratio against [Fe/H] in unmixed stars ($\log
L/\Lsol<1.5$). Different symbols are used for C abundances obtained from
different indices} 
\label{fig:7}
\end{figure}

We may now plot in Figure~\ref{fig:7} the run of the [C/Fe] ratio with [Fe/H]
for {\it bona fide} unmixed stars (i.e. those with $\log L/\Lsol<1.5$).
Figure~\ref{fig:7} shows that the [C/Fe] ratio is roughly solar over the whole
metallicity regime explored in this paper. The moderate overabundances in the
most metal-poor disk stars suggested by Andersson and Edvardsson (1994) and
Tomkin et al (1995) are reduced to a marginally significant $\sim 0.1$~dex
value in our reanalysis of their data: this is mainly due to the adoption of a
higher \teff\ scale. We finally warn the reader that the [C/Fe] run of
Figure~\ref{fig:7} is obtained by combining data provided by the analysis of
various features; while the overall trend seem well defined (different indices
yielding similar results), small offsets might still be present. A more
comprehensive study overcoming the selection biases present in all
available samples would be welcome.

\begin{figure}
\psfig{figure=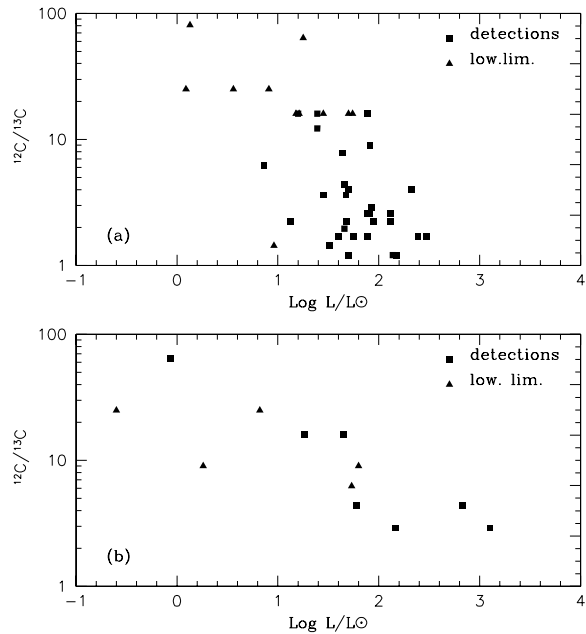,width=10.0cm,clip=}
\caption[ ]{Run of the \C12C13 ratio against stellar luminosity for stars with
$-0.9<$[Fe/H]$<-0.2$\ (panel a), and for stars with [Fe/H]$<-0.9$\ (panel b).
Plotted data includes \C12C13 ratios determined in this paper, as well as
those by Sneden et al. (1986) and Shetrone et al. (1993). Rectangles represent
actual detections, while triangles represent lower limits} 
\label{fig:8}
\end{figure}

\subsection{\C12C13 }

Panels a and b of Figure~\ref{fig:8} display the run of the \C12C13 ratio
against stellar luminosity for stars with $-0.9<$[Fe/H]$<-0.2$, and for stars
with [Fe/H]$<-0.9$\ respectively. 

Plotted data include \C12C13 ratios determined in this paper, as well as
those by Sneden et al. (1986) and Shetrone et al. (1993). Both panels indicate
that the \C12C13 ratio declines with increasing luminosity: in fact, stars with
$\log L/\Lsol<1.5$\ have very large values of the \C12C13 ratios (in most
cases, only lower limits can be determined), while very small values (4--10)
have been obtained for stars with $\log L/\Lsol>2$. These data confirm earlier
findings for both field and cluster stars (see e.g. Gilroy \& Brown 1991; and
Gratton et al. 1999b), and the lack of evidences for mixing in stars with
$\log L/\Lsol<1.5$.

\subsection{Nitrogen }

N abundances are listed in Columns 5, 6 and 7 of Table~\ref{tab:3}; they were
obtained only for stars in our original sample. In some spectra, CN features
are very weak, and only upper limits can be obtained. Furthermore, there might
be a systematic offset in our nitrogen abundances, due both to errors in the C
abundances and to uncertainties in the same analysis of CN (e.g. related to
uncertainties in the dissociation potential for this molecule). Conclusions are
thus not as clean as those obtained for the other elements. The comparison
between abundances provided by lines due to 4 blends of the blue system
(between 4211 and 4216~\AA) and about 40 blends of the red system (between 7912
and 7980~\AA) is rather good. On average, N abundances given by the blue system
are larger by $0.12\pm 0.07$~dex ($\sigma =0.19$~dex, 8 stars). A large part of
this difference is due to a single star (HD~136316) having very weak lines of
the red system; once this star is dropped, the average difference is $0.07\pm
0.04$~dex ($\sigma =0.12$~dex, 7 stars), which may be attributed to small
errors in the location of the continuum level in our line crowded, blue
spectra. In the following, we simply averaged the N abundances provided by the
two systems.

\begin{figure}
\psfig{figure=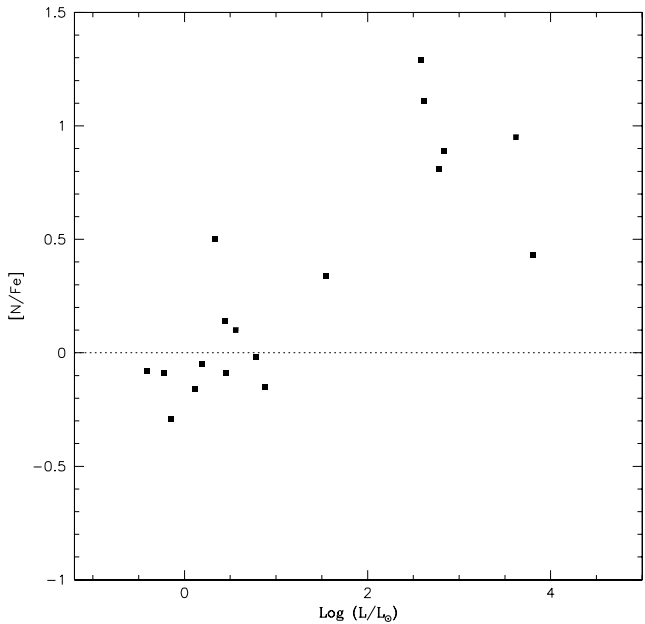,width=9.0cm,clip=}
\caption[ ]{Run of the [N/Fe] ratios against stellar luminosity for our 
program stars} 
\label{fig:9}
\end{figure}

The run of the [N/Fe] ratio against stellar luminosity is shown in
Figure~\ref{fig:9}. There is an obvious trend for increasing N abundances with
increasing luminosity which is clearly symmetric with respect to the run of
the [C/Fe] ratio: stars with $\log L/\Lsol<2$\ have [N/Fe]=$+0.02\pm 0.07$~dex
($\sigma =0.22$~dex, 11 stars), while more luminous stars have [N/Fe]=$+0.91\pm
0.11$~dex ($\sigma =0.27$~dex, 6 stars). As well known, this is exactly what
expected if N-rich, C-poor material is brought to the surface from regions of
CNO-cycle H-burning.

Mean [(C+N)/Fe] ratios (Table~\ref{tab:3}) for stars fainter and brighter than
$\log L/\Lsol=2$\ are $0.03\pm 0.05$~dex ($\sigma =0.17$~dex, 13 stars) and
$0.32\pm 0.08$~dex ($\sigma =0.19$~dex, 6 stars) respectively. However, we are
not inclined to give much weight to this difference, due to possible systematic
errors in abundances deduced from molecular species related to the uncertain
values of the dissociation energy. Moreover, in our sample, the most luminous
stars are also more metal-poor: hence, the systematic trend of the [(C+N)/Fe]
ratio with luminosity might be due to a trend with metal abundance.

Unluckily, our sample cannot be used to constrain the run of the [N/Fe] ratio
with metal abundance, since N abundances were obtained only for metal-rich
stars or stars in which mixing events have already changed the original
abundances. 

\subsection{Sodium }

Non-LTE corrections for the Na lines considered in the present work are rather
small ($<0.08$~dex) both for the Sun and for moderately metal-poor dwarfs (in
these last cases they are slightly positive: $\sim 0.03$~dex). However, since
corrections depend on stellar gravity, they are not negligible (up to 0.3~dex)
for metal-poor giants. Moreover, corrections for these stars depend on the
adopted value for the collisional cross-section (see also Paper II for a
detailed discussion). In order to quantify this dependence, we repeated our
statistical equilibrium computations for a typical metal-poor giant
(HD~122956), using different values for the multiplicative constant $x$ for the
collisional cross-sections: $x=$0.001, 0.01 and 0.1. As expected, non-LTE
corrections are negligible for large $x$ values, but they are rather large
(about 0.4~dex) for $x$=0.001. This last value yields cross sections much
smaller than those estimated by Kaulakys (1985, 1986), but it is however still
compatible with RR Lyrae observations (see also Paper II). 

\begin{table}
\caption[ ]{Na abundances in the 19 program stars (available only in 
electronic form)}
\label{tab:8}
\end{table}

Table~\ref{tab:8} (available only in electronic form) 
lists sodium abundances (with and without non-LTE corrections), along with
the $EW$s of individual lines for the stars we observed directly. The average
[Na/Fe] ratio for stars with [Fe/H]$<-0.6$ is $-0.09\pm0.05$ ($\sigma=0.19$, 13
stars). The scatter decreases if we omit the two coolest stars: in this case,
the average is $-0.04\pm 0.04$ ($\sigma=0.13$, 11 stars). The agreement
obtained from different lines is excellent ($\sigma=0.08$~dex). 

\begin{figure}
\psfig{figure=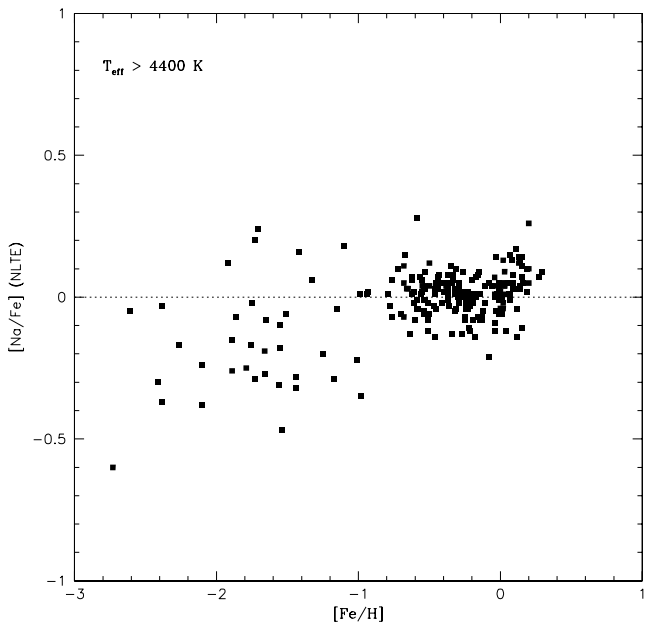,width=9.0cm,clip=}
\caption[ ]{Run of the [Na/Fe] ratio against overall metal abundance [Fe/H] 
for the stars of the total sample analyzed in the present paper} 
\label{fig:10}
\end{figure}

Figure~\ref{fig:10} displays the run of the [Na/Fe] ratio with overall metal
abundance for all stars considered in this paper; sodium abundances scale 
approximately as iron ones, being possibly overdeficient in stars with
[Fe/H]$\sim -1$. The scatter is larger than expected from the line-to-line
comparison; it may be due to uncertainties in the atmospheric parameters, which
might induce rather large errors in the [Na/Fe] ratios. However, part of this
scatter might be real, since sodium is known to be overabundant in some
globular cluster giants (see Kraft et al. 1992; Carretta \& Gratton 2000). 

\subsection{Magnesium } 

Barbuy et al. (1985) and Magain (1989) suggested that large discrepances exist
between
abundances provided by the resonance intercombination line and higher
excitation Mg lines in the spectra of red giants and subdwarfs.
Magain estimated differences as large as 0.6 dex, apparently a strong function
of surface gravity; he suggested that the dominant non-LTE effect for Mg should
be overionization, and that the resonance line may require larger non-LTE
corrections since there is a larger contribution in its formation from
the outer atmospheric layers than in the case of the higher excitation
lines. However, other studies obtained a much better agreement between
abundances provided by different Mg lines: in an analysis of the spectra of RR
Lyrae variables (Clementini et al. 1995), it was found that the Mg abundances
deduced from the intercombination resonance line at 4571~\AA\ are larger than
those provided by higher excitation lines by $0.09\pm 0.08$~dex; a similar
result has been obtained in the present analysis for field, not-variable stars
(in this case the mean difference is $0.06\pm 0.05$~dex). This is due both to
the different value for the $gf$\ of the intercombination line (see
Section~3.6), and to the adoption of a higher \teff-scale; however, to better
clarify this issue we decided to perform detailed statistical equilibrium
computations for a Mg~I model atom (see Paper II for details). 

\begin{table}
\caption[ ]{Magnesium abundances in the 19 program stars
(available only in electronic form) }
\label{tab:9}
\end{table}

The results of our non-LTE analysis for Mg are listed in Table~\ref{tab:9}
(available only in electronic form) for
the 19 stars of the original sample; both the abundances given by individual
features and the mean Mg values are indicated. The line-to-line scatter is
moderately large ($\sigma =0.13$~dex), due to the rather large range in
excitation potential and line strength. The Mg to Fe ratio is moreover rather
sensitive on the adopted values for the effective temperature and
microturbulent velocity. On the whole, derived abundances are very similar to
those (not shown here) obtained under LTE assumption: corrections are
$<0.02$~dex for the observed high excitation lines, that form at large optical
depth and are usually very weak. Very small non-LTE corrections are found even
for the intercombination line; they are in the range 0.05--0.10~dex only for
cool and moderately metal-poor giants (HD~136316 and HD~187111). In these stars
this line is very strong and heavily saturated. 

In order to check the sensitivity of our statistical equilibrium computations
to the value of the parameter $x$, the non-LTE abundance analysis has been
repeated for the star having the largest corrections ($i.e.$ HD~187111). For
this purpose, we used $x=$7, at the lower edge of the fiducial range as deduced
from the observation of RR Lyrae stars (Paper II). As expected, our results
show that Mg abundances are about 0.05~dex larger for all lines. We can then
conclude that upper limits to non-LTE corrections for Mg abundances are
0.15~dex for the intercombination line and 0.10~dex for the other lines, in the
stars of our program sample. The low Mg abundances obtained by Barbuy et al.
(1985) and Magain (1989) when using the 4571~\AA\ line cannot be explained as
due to departures from LTE, and must be rather attributed to choice of 
the oscillator strength. 

The [Mg/Fe] ratios of our program star sample seem to follow without
discontinuity the run with overall metal abundance [Fe/H] obtained from the
reanalysis of the E93 sample (listed in Table~\ref{tab:6}). However, some 
discrepancy is
present in the abundances obtained from the reanalysis of the ZM90 sample (see
Table~\ref{tab:7}). In this case [Mg/Fe] ratios derived from the
intercombination line well agree with those obtained for our 19 program stars:
mean values for stars with [Fe/H]$<-0.6$ are $+0.40\pm 0.04$~dex and $+0.38\pm
0.04$~dex respectively, once the appropriate corrections for deviations from
LTE (that are little, anyway) are applied. On the other side, [Mg/Fe] ratios
derived from high excitation lines are lower than those obtained for our
sample: $0.27\pm 0.03$ as compared to $0.38\pm 0.04$. 

\begin{figure}
\psfig{figure=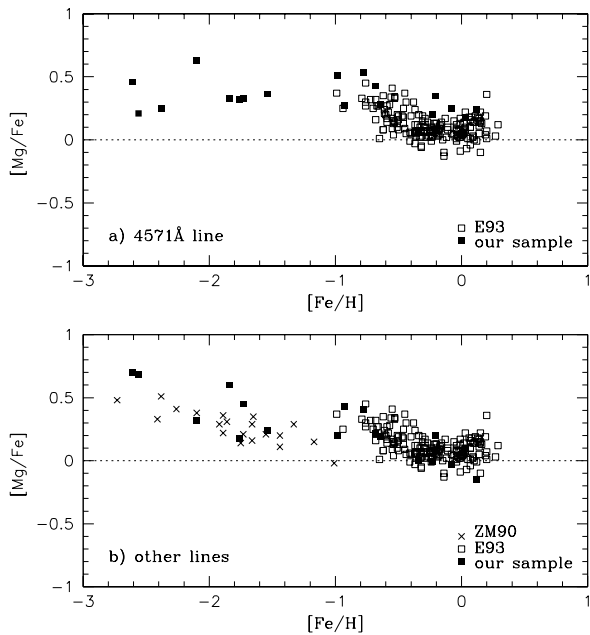,width=10.0cm,clip=}
\caption[ ]{Run of the [Mg/Fe] ratio against overall metal abundance [Fe/H].
In panel a) are plotted results from the intercombination line at 4571~\AA\ and
in panel b) those derived from higher excitation lines, for all the stars in
our total combined sample} 
\label{fig:11}
\end{figure}

Figure~\ref{fig:11} displays the run of the [Mg/Fe] ratio with overall metal
abundance [Fe/H] from the intercombination line and the higher excitation
lines. The reason of this discrepancy is not completely clear; we examined
different hypoteses, like systematic differences in the oscillator strengths, in
$EWs$ and in adopted values of the microturbulent velocity. Our conclusion is
that none of these causes can explain the observed difference. More promising
is the possibility that departures from LTE (computed for the ZM90 sample
with the $x$ parameter recommended from the analysis of RR Lyrae stars) could
be underestimated. In fact, corrections derived from the high excitation
lines are larger for the warmest and most metal-poor stars, as the stars in the
ZM90 sample are: while corrections for the intercombination line are larger in
the coolest metal-poor stars. 

Notwithstanding this (small) discrepancy, in the discussion of Paper IV we will
consider the mean values of [Mg/Fe] obtained using all lines. 

\section{Comparisons with other recent abundance determinations}

\begin{figure}
\psfig{figure=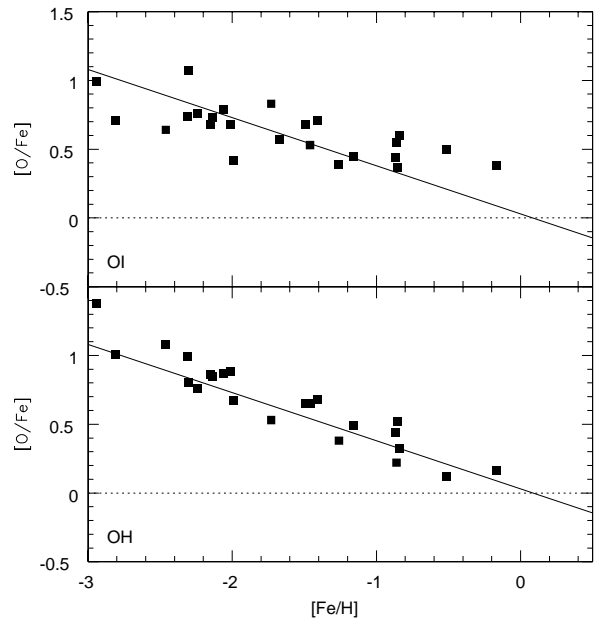,width=10.0cm,clip=}
\caption[ ]{Run of the [O/Fe] ratio with [Fe/H] determined by Boesgaard et al.
(1999) using OI lines (upper panel) and the UV OH band (lower panel).
Overimposed is the mean relation 
[O/Fe]=$-0.35~(\pm 0.03)$~[Fe/H]~+$0.03~(\pm 0.05)$\ proposed by Boesgaard
et al.} 
\label{fig:12}
\end{figure}

\subsection{Boesgaard et al. 1999}

Very recently, Boesgaard et al. (1999) published an analysis of the O abundances
from OH lines, and reanalyzed literature data concerning OI (including some
of the same data here considered, that is TLLS). Boesgaard et al. gave two
different sets of abundances, derived using model atmospheres which used
two different temperature scales. The highest one (labeled "King") agrees very
well with the present one: for 20 stars in common, the mean difference is
only $6\pm 15$~K, with a scatter of 65~K for individual stars. Gravities are
derived by averaging values from ionization equilibrium (the technique used in
this paper) with those given by a calibration of Str\"omgren photometry. These
gravities also compare very well with ours, the mean difference being $0.14\pm
0.03$~dex, with a scatter for each star of 0.27~dex. Finally, their
microturbulent velocities are on average $0.45\pm 0.08$~km~s$^{-1}$\ larger
than ours: these larger values of the microturbulent velocity explain about
half the average difference in the [Fe/H]'s (their values being smaller than
ours on average by $0.09\pm 0.02$~dex). Boesgaard et al. O abundances from
permitted atomic lines are on average larger than ours by $0.09\pm 0.03$~dex
(scatter 0.13~dex), a difference that may be attributed to our adopted
corrections for departures from LTE (these corrections are not included in the
Boesgaard et al.'s O abundances). While we regard these differences as minor,
we note that combining the higher O abundances with the lower [Fe/H]'s, on
average the [O/Fe] values of Boesgaard et al. are higher than ours by 0.18~dex.

Since the present analysis reproduces quite well the results by Boesgaard et
al., apart from small offsets in [Fe/H] and [O/Fe] that may be easily
explained, it seems quite odd that our Figure~\ref{fig:4} does not support
well their mean relation [O/Fe]=$-0.35~(\pm 0.03)$~[Fe/H]~+$0.03~(\pm 0.05)$\
(a similar steep slope was previously obtained by Israelian et al. 1998, still 
from
analysis of OH bands in the extreme ground-based UV).
Rather, our O abundances suggest a mild slope ($\sim -0.1$) in the [O/Fe]
ratios for stars with [Fe/H]$<-1$, with a rather abrupt change in the range
$-1<$[Fe/H]$<-0.5$, and then an almost constant value for higher metallicities
(in Paper IV we will interprete such a relation within the framework of
galactic evolution). Part of this difference may be attributed simply to
statistics: we considered a much larger number of stars, mainly in the
critical metal abundance range $-1<$[Fe/H]$<-0.5$, corresponding to the
transition from thick to thin disk. However, we note that the constant slope
of the run of average [O/Fe] vs [Fe/H] found by Boesgaard et al. is due
to their O abundances from OH, and not to those from OI lines. To show this,
we plotted in Figure~\ref{fig:12} the runs of the [O/Fe] values with [Fe/H]
obtained by Boesgaard et al. with O abundances from OI lines and from OH.
Overimposed is the Boesgaard et al. mean relation: while this provides an
excellent fit to the abundances from OH, the slope suggested by OI lines is
much milder, and roughly in agreement with the impression given by our
Figure~\ref{fig:4}. While the issue of O abundance determinations is clearly
not yet definitely settled, we argue that abundances from OH for stars in the
metal abundance range $-1<$[Fe/H]$<-0.5$\ may be regarded with suspicion,
since location of the continuum level is difficult at these wavelengths in
such line rich spectra.

\subsection{Gratton et al. 1999b}

We are now publishing a separate paper on mixing episodes along the RGB
(Gratton et al. 1999b). The two analysis are almost completely independent each
other, using different observational material, a different set of atmospheric
parameters, different line lists, and even slightly different model
atmospheres (those used here are the Kurucz models with the overshooting
option switched on; the paper on mixing uses the model atmospheres
with the overshooting option switched off). However, there are 5 stars in
common between the two samples (here we only considered stars in the original
samples); they are all giants, with somewhat uncertain reddening estimates. On
average adopted atmospheric parameters are quite similar, although with a
quite large star-to-star scatter: present $T_{\rm eff}$'s and gravities are
lower by $73\pm 73$~K ($\sigma=164$~K), and $0.19\pm 0.33$~dex
($\sigma=0.74$~dex), respectively. Average abundances are also quite similar:
differences (in the sense this paper-Gratton et al. 1999b) are are $-0.03\pm
0.09$, $0.13\pm 0.07$, $0.00\pm 0.08$, and $-0.10\pm 0.23$~dex for [Fe/H],
[C/Fe] (from CH), and [O/Fe] (from [OI] and OI lines), respectively. Scatter
for individual stars is however quite large (0.21, 0.15, 0.17, and 0.33 dex).
These slightly different results are due to the different choice about the
atmospheric parameters. If we would adopt the same parameters for these stars
used in Gratton et al. (1999b), we would have obtained average differences of
$0.07\pm 0.02$, $0.03\pm 0.06$, $0.06\pm 0.04$, and $-0.03\pm 0.02$~dex for
[Fe/H], [C/Fe] (from CH), and [O/Fe] (from [OI] and OI lines), respectively,
with much smaller scatter individual stars (0.05, 0.14, 0.10, and 0.03 dex).
Finally, for Na we compared the abundances of this paper with those for the
extended sample of Gratton et al. (1999b): from 21 stars, we get an average
offset of $0.01\pm 0.05$~dex ($\sigma =0.22$~dex).

\subsection{Fuhrmann 1998}

A further interesting comparison is with the very precise recent analysis
of nearby dwarfs and subgiants by Fuhrmann (1998). There are 32 stars in
common between our paper and that of Fuhrmann. The agreement is very good:
on average, our [Fe/H] are smaller by $0.02\pm 0.01$~dex ($\sigma=0.08$~dex),
while the [Mg/Fe] values are larger by $0.01\pm 0.02$~dex ($\sigma=0.10$~dex).
In Paper IV, when we will discuss the implications of the present data for
galactic evolution, we will see that the conclusions we reach are very similar 
to those of Fuhrmann.

\section{ Conclusions }

In this paper we have presented an analysis of the abundances of Fe, C, N, O,
Na, and Mg in a large sample ($\sim 300$) of field stars. Original data
consisted in a new set of high resolution, high S/N spectra for 19 stars; these
were complemented with high quality $EW$s from selected literature surveys. 
Whenever possible, we tried to compare results obtained with several abundance
indices.

The main conclusions are: 
\begin{enumerate}
\item Once the (small) corrections for non-LTE effects (presented in
Gratton et al. 1999: Paper II) are considered, O abundances derived from the
permitted lines are fully consistent with those obtained using the forbidden
line at 6300.3~\AA\ for stars with \teff$>$4600~K; this result agrees with that
obtained by King (1993), but not with other studies using a lower \teff-scale
for subdwarfs. The O overabundance in metal-poor stars ([Fe/H]$<-0.8$) are
$0.48\pm 0.05$~dex ($\sigma=0.16$~dex, 11 stars) and $0.45\pm 0.02$~dex
($\sigma=0.13$~dex, 33 stars) from forbidden and permitted lines respectively;
the mean value is $0.46\pm 0.02$~dex ($\sigma = 0.12$, 32 stars). However, the
most luminous metal-poor field stars yield slightly smaller O abundances than
fainter stars of similar metallicities. This is attributed to inadequacy of
Kurucz (1992) model atmospheres with overshooting for stars cooler than $\sim
4600$~K. Our data suggest a general overabundance of O in metal-poor stars
([Fe/H]$<-1$), with only a gentle trend for decreasing O excesses with
increasing metal abundances; these results are not well fitted by a constant
slope in the relation between [O/Fe] and [Fe/H] recently suggested from the
analysis of OH lines.
\item If only {\it bonafide} unmixed stars are considered, C abundances
scale as Fe ones over the whole range explored ([Fe/H]$<-2.5$). We did not 
confirm the moderate C excess found by Andersson and Edvardsson (1994) and
Tomkin et al (1995) in metal-poor disk stars ($-0.8<$[Fe/H]$<-0.4$); this is
due to our adoption of a higher \teff\ scale. 
\item Na abundances scale as Fe ones in the high metallicity regime, while
metal-poor stars present a Na underabundance. None of the field stars analyzed
here belong to the group of O-poor and Na-rich stars observed in globular
clusters. 
\end{enumerate}

In future papers of this series we will discuss these results in the framework
of galactic chemical evolution. 


%
%
%
%
%
%
%

\end{document}